\documentclass[aps,twocolumn,groupedaddress,amssymb,prl,floatfix,preprintnumbers]{revtex4}
\usepackage{graphicx}

 \newcommand{\etal}      {{\it et al}.}

 \newcommand{\dpu}       {$\delta$-Pu}
 \newcommand{\pu}       {$^{239}$Pu}
 \newcommand{\slrrtext}  {spin lattice relaxation rate}
 \newcommand{\slrr}      {$T_1^{-1}$}

\begin{document}
\bibliographystyle{apsrev}
\preprint{LA-UR-03-8735}

\pagestyle{empty}

\title{Nuclear Magnetic Resonance Studies of $\delta$-Stabilized Plutonium}
\author{N. J. Curro}
\affiliation{Condensed Matter and Thermal Physics, Los Alamos
National Laboratory, Los Alamos, NM 87545,
\texttt{curro@lanl.gov}}
\author{L. Morales}
\affiliation{Nuclear Materials and Technology Division, Los Alamos
National Laboratory, Los Alamos, NM 87545}

\date{\today}

\begin{abstract}

  Nuclear Magnetic Resonance studies of Ga stabilized \dpu\ reveal
  detailed information about the local distortions surrounding the
  Ga impurities as well as provides information about the local
  spin fluctuations experienced by the Ga nuclei.  The Ga NMR
  spectrum is inhomogeneously broadened by a distribution of
  local electric field gradients (EFGs), which indicates that the
  Ga experiences local distortions from cubic symmetry.  The
  Knight shift and \slrrtext\ indicate that the Ga is dominantly
  coupled to the Fermi surface via core polarization, and is {\it
  inconsistent} with magnetic order or low frequency spin
  correlations.

\end{abstract}

\maketitle

\subsection{Introduction}

The investigavtion of the low temperature properties of plutonium
and its compounds has experienced a renaissance in recent years,
and several important experiments have revealed unusual correlated
electron behavior
\cite{sarraoPuCoGa5nature,SmithBoringLANL,LashleySpecificHeat}.
The 5f electrons in elemental plutonium are on the boundary
between localized and itinerant behavior, so that slight
perturbations in the Pu-Pu spacing can give rise to dramatic
changes in the ground state character.  In fact, Pu exhibits six
allotropic forms in the solid state,  with varying degrees of
symmetry.  The inability for band structure calculations to
predict all six of these forms is testament to the complexity of
the correlated electron behavior \cite{wills, sarasov}. \dpu,
which exists between 576K and 720K, is the most technologically
important form, and can be stabilized down to $T=0$ by doping with
Al, Ga or In, which substitute for the Pu sites \cite{hecker}. In
particular, Pu$_{1-x}$Ga$_x$ possesses the fcc structure and
physical properties of \dpu\ for $0.020\lesssim x \lesssim 0.085$.
It is not obvious why the small amount of the secondary element
will stabilize the electronic structure of the $\delta$ phase,
although clearly the slight perturbation reflects the
itinerant-localization instability of the parent material.

Several theories have emerged to try to explain the stability of
the $\delta$ phase.  One technological challenge of band
calculations in Pu is the ability to predict local versus
itinerant behavior for the five 5f electrons, because the
theoretical approaches for these two extremes are fundamentally
different.    These models typically predict either non-magnetic
ground states, or ones where local Pu moments are retained
\cite{wills, magneticPutheory}.   In principle, such predictions
can be tested by various experimental techniques.

Nuclear magnetic resonance (NMR) is ideal for probing magnetic
correlations and local structure. \pu\ has a nuclear spin
$I=\frac{1}{2}$, so in principle it can serve as a direct probe of
the magnetic correlations via the hyperfine interaction
$\mathbf{A}$  between the nuclear ($I$) and the electronic ($S$)
spins.  However, for 4f and 5f nuclei, $\mathbf{A}$ is usually
quite large (on the order of 1000 kOe/$\mu_B$), so that
fluctuations of of the electron spin $S$ relax the nuclei so fast
that their signal is rendered invisible.  On the other hand, the
hyperfine coupling to the nuclei of the secondary element (Al, Ga
or In) are typically one to three orders of magnitude smaller, so
by measuring the secondary nuclei one can gain considerable
insight into the spin dynamics of the system.

One of the challenges facing band theorists calculating the
electronic structure of \dpu\ is the role of the secondary atoms.
Recent x-ray absorbtion fine structure (XAFS) studies suggest that
Pu lattice distorts locally around the Ga sites \cite{conradson}.
Clearly, these local distortions are caused by the electronic
system, and should be captured by band structure calculations.
Therefore, a detailed experimental study of these local
distortions puts important constraints on any theory. NMR is also
ideal for probing these local distortions, and serves as a
complement to the XAFS data. Nuclei with spin $I>\frac{1}{2}$
experience electric quadrupolar interactions between the
quadrupolar moment $Q$ of the nuclei and the local electric field
gradient (EFG), $V_{\alpha\alpha}$, where $V$ is the crystal field
potential.  Ga ($I=\frac{3}{2}$) is therefore sensitive to the EFG
at the impurity site.

In this paper we present Knight shift ($K$), \slrrtext\ (\slrr),
and linewidth ($\sigma$) data on the Ga nucleus in a polycrystal
sample of Pu$_{1-x}$Ga$_x$ for $x=0.017$  between 4K and 100K. In
the early 1970's, Fradin, Brodsky, and coworkers at Argonne
performed the first NMR experiments on \dpu\ stabilized with Al
\cite{Fradin1970}. They found no evidence for any magnetic
transition or the formation of local moments. Our work on Ga
stabilized \dpu\ provide the only new data since the pioneering
work of Fradin, and we find similar behavior. The Knight shift and
\slrr\ data behave as in a typical conductor, and we find no
evidence for either local moments at the Ga sites, or for strongly
temperature dependent spin fluctuations \cite{fluss}. On the other
hand, we do not find that the EFG at the Ga site vanishes, as
expected for a site with cubic symmetry. Rather, the Ga spectrum
is inhomogeneously broadened by a distribution of EFGs. This
result suggests that there are local lattice distortions with a
symmetry lower than cubic at the Ga impurities.

\subsection{Experimental}

In order to avoid radiological contamination of the NMR probe, we
constructed an epoxy assembly to encapsulate the NMR coil as well
as the Pu sample.  However, since the Pu is subject to both
self-heating as well as rf heating due to induced surface currents
by the NMR pulses, it is crucial to establish a solid thermal
contact to the sample. We therefore bored out the epoxy along the
axis of the NMR coil, and secured frits (metallic screens) at the
ends to allow the cold He gas to enter along the axis of the coil
and establish a thermal contact with sample. The mesh of the frits
is sufficiently fine to prevent any external contamination. This
design allows us to cool the sample down to 2K with no significant
heating problems, as measured by the nuclear polarization.

\begin{figure}
  \centering
  \includegraphics[width=\linewidth]{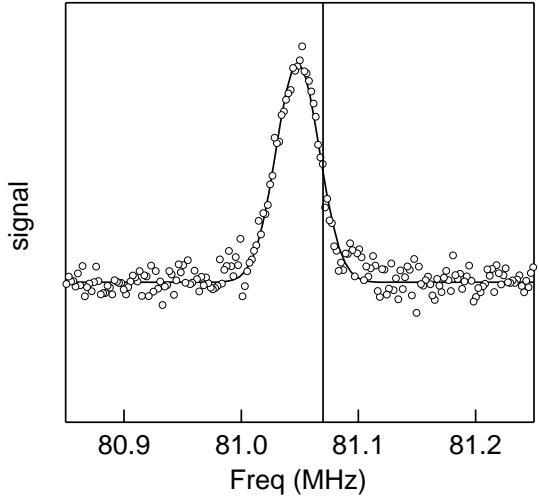}
  \caption{The $^{69}$Ga NMR spectrum at
  4K.
  The solid vertical line is the position of pure Ga metal, and the
  solid line through the data is a Gaussian fit as described in the text.}
  \label{fig:spectra}
  \end{figure}

The Hamiltonian of the Ga nuclei is given by:
\begin{equation}
\mathcal{H}=\gamma\hbar\hat{I}\cdot\mathbf{H}_0+\frac{h\nu_c}{6}(3\hat{I}_z^2-I^2+\eta(\hat{I}_x^2-\hat{I}_y^2))+\mathcal{H}_{\rm
hyp}
\label{eqn:NMReqn}
\end{equation}
where $\gamma$ is the gyromagnetic ratio, $\mathbf{H}_0$ is the
external field, $\nu_{c} = 3eQV_{cc}/20$,
$\eta=(V_{aa}-V_{bb})/V_{cc}$, $Q$ is the quadrupolar moment of
the Ga, and $V_{\alpha\alpha}$ are the components of the EFG
tensor. The hyperfine interaction is given by $\mathcal{H}_{\rm
hyp} =
\mathbf{I}\cdot\sum_i\mathbf{A}_i\cdot\mathbf{S}_i(\mathbf{r})$,
where the sum is over the various spin contributions (conduction
electrons, local spins).  For a site with cubic symmetry,
$V_{\alpha\alpha}$ vanishes, so the effective Hamiltonian for the
Ga nuclei in \dpu\ is simply:
\begin{equation}
\mathcal{H}=\gamma\hbar(1+K)\hat{I}_zH_0,
 \label{eqn:NMRKnight}
\end{equation}
where
\begin{equation}
K(T)=K_0 + \sum_iA_i\chi_i(T),
 \label{eqn:Knight}
\end{equation}
$K_0$ is a temperature independent orbital shift contribution, and
$A_i$ is the hyperfine coupling to the $i^{th}$ component of the
electronic susceptibility, $\chi(T)$.

\begin{figure}
  \centering
  \includegraphics[width=\linewidth]{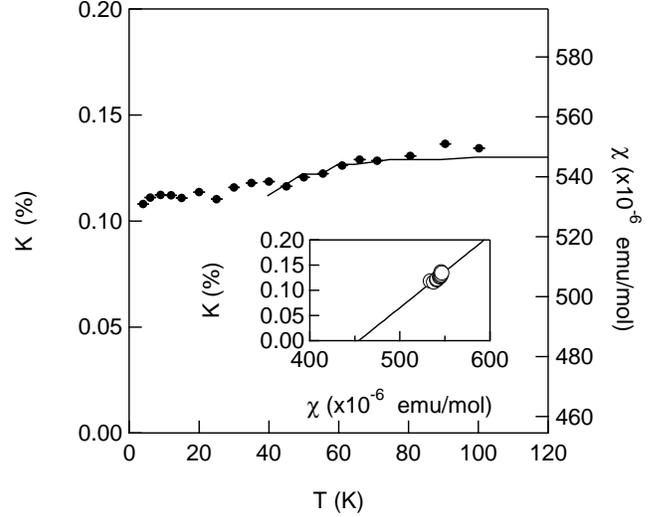}
  \caption{The Knight shift ($\bullet$) and susceptibility (solid line) of the Ga versus
  temperature. The susceptibility data are taken from
  \cite{fournier}.
   INSET: The Knight shift versus bulk susceptibility.}
  \label{fig:Knight}
  \end{figure}

\begin{figure}
  \centering
  \includegraphics[width=\linewidth]{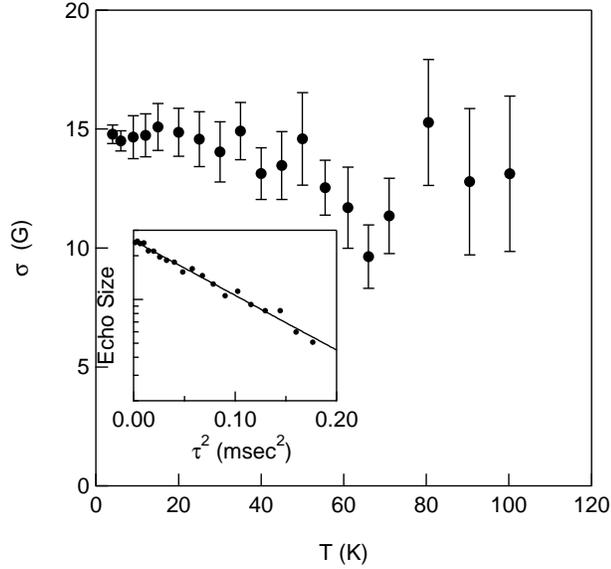}
  \caption{The rms second moment of the spectral linewidth versus temperature of the $^{69}$Ga line in \dpu. INSET:
  The echo integral versus $\tau^2$, where $\tau$ is the pulse spacing between the 90$^\circ$ and $180^\circ$ pulses.}
  \label{fig:FWHM}
  \end{figure}

In Fig. (\ref{fig:spectra}) we show the Ga spectrum at 4K.  Note
that this spectrum consists of all three nuclear transitions
($I_z=+\frac{3}{2}\leftrightarrow+\frac{1}{2},I_z=+\frac{1}{2}
\leftrightarrow-\frac{1}{2},I_z=-\frac{1}{2}\leftrightarrow-\frac{3}{2}$),
since the quadrupolar Hamiltonian does not lift the degeneracy.
The spectrum is broadened, however, by a distribution of EFG's, as
we discuss below.  The center of the resonance is somewhat lower
in frequency than that of pure Ga metal, indicating a different
Knight shift. We fit the spectra between 4K and 100K to a
Gaussian; $K$ and the rms second moment, $\sigma$, are shown in
Figs. (\ref{fig:Knight},\ref{fig:FWHM}).  If we assume a single
component of magnetic susceptibility, then following Eq.
(\ref{eqn:Knight}), we extract $K_0$ and $A$ by plotting $K$
versus $\chi$, where $T$ is an implicit parameter (see inset of
Fig. (\ref{fig:Knight}). We find $K_0=0.65\%$ and
$A=80$kOe/$\mu_B$.

  \subsection{Spin Lattice Relaxation}

The spin lattice relaxation measures the time scale for the
nuclear spin system to acquire an equilibrium Boltzmann
distribution among the energy levels.  Transitions between the
nuclear levels are induced by time dependent fluctuating fields
that are perpendicular to $\mathbf{H}_0$.  The spin lattice
relaxation rate, \slrr\ is give by:
\begin{equation}
\frac{1}{T_1T} = \frac{\gamma^2k_B}{2}\lim_{\omega\rightarrow 0}
\sum_{\mathbf{q}}A_{\perp}^2(\mathbf{q})
\frac{\chi_{\perp}"(\mathbf{q},\omega)}{\omega} \label{eqn:T1}
\end{equation}
where $A_{\perp}(\mathbf{q})$ is the spatial Fourier transform of
the hyperfine coupling in the perpendicular direction,
$\chi_{\perp}"(\mathbf{q},\omega)$ is the dynamical
$\mathbf{q}$-dependent susceptibility, and the sum is over the
Brillouin zone. In this case, we expect that $A$ is isotropic and
$\mathbf{q}$-independent, thus the \slrrtext\ is sensitive to
fluctuations for all $\mathbf{q}$, and any magnetic correlations
or magnetic order should be reflected in \slrr.

\begin{figure}
  \centering
  \includegraphics[width=\linewidth]{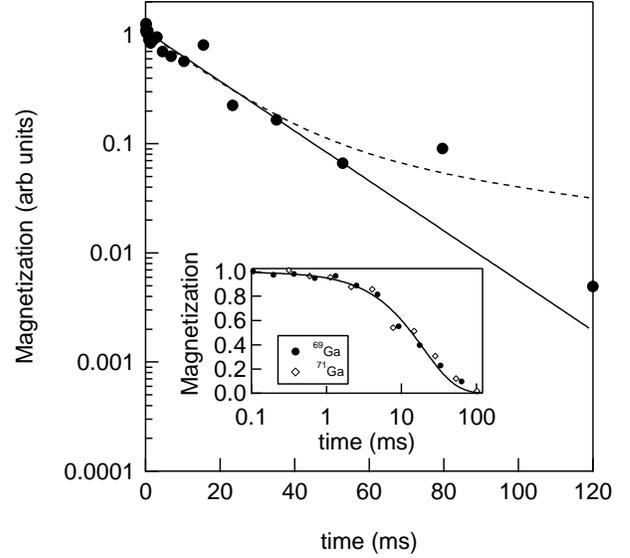}
  \caption{The magnetization decay of the Ga line at 4K, and fits
  to a single exponential decay (solid line) and the
   decay function for the central transition (dotted line). INSET: The magnetization recovery data for
   both isotopes of Ga.  The time scale for the $^{71}$Ga has been scaled by $(^{71}\gamma/^{69}\gamma)^2$, as
   expected for magnetic relaxation. }
  \label{fig:magrecover}
  \end{figure}

We measured  the $^{69}$Ga \slrr\  by inversion recovery. In Fig.
(\ref{fig:magrecover}) we show data at 4K, which shows a single
exponential.  In principle, the recovery of the $I=\frac{3}{2}$ Ga
can be more complex.  However, since the satellites overlay the
central transition (strictly speaking, the $H_1$ pulses have a
bandwidth much greater than the EFG distribution) all nuclear
levels are inverted by the pulses, and the recovery function
simplifies to a single exponential of the form:
\begin{equation}
M(t) = M_0(1-fe^{-t/T_1}),
\end{equation}
where $M_0$, $f$, and $T_1$ are fit parameters. If the EFG were
actually much greater, and the satellite transitions
($I_z=\pm\frac{3}{2}\leftrightarrow\pm\frac{1}{2}$) were either
too broad to observe, or missing from the spectra for some reason,
then the recovery of the observed central transition
($I_z=+\frac{1}{2}\leftrightarrow-\frac{1}{2}$) would be of the
form:
\begin{equation}
M(t) =
M_0\left[1-f\left(\frac{9}{10}e^{-6t/T_1}+\frac{1}{10}e^{-t/T_1}\right)\right].
\end{equation}
Clearly, as observed in Fig. (\ref{fig:magrecover}), the data
indicate a single exponential, which supports the conclusion that
the Ga is located at cubic site.  In the inset, we compare the
relaxation of the $^{71}$Ga with that of the $^{69}$Ga.  The time
scale for the $^{71}$Ga has been scaled by
$(^{71}\gamma/^{69}\gamma)^2$ (see Eq. (\ref{eqn:T1})); the fact
that the data for both sets of isotopes fall onto the same line
indicates a magnetic relaxation mechanism, rather than a
quadrupolar one (as might be expected for structural fluctuations,
for example).

\begin{figure}
  \centering
  \includegraphics[width=\linewidth]{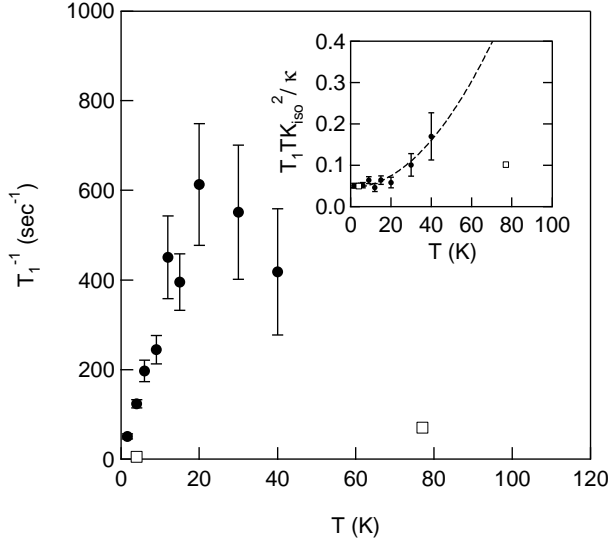}
  \caption{The \slrrtext\ versus temperature for the Ga ($\bullet$), and
  Al ($\square$, from \cite{fradin}) in \dpu.  INSET: $T_1TK^2/\kappa$ versus $T$, where $\kappa$ is
  given by the Korringa constant. The  dotted line is a fit to $T_1TK^2/\kappa = S_0(1+(T/T_0)^2)$,
   where $S_0=0.046$, and $T_0=25$K.  }
  \label{fig:T1Tdep}
  \end{figure}

In Fig. (\ref{fig:T1Tdep}) we show the temperature dependence of
\slrr.  In conductors where the dominant relaxation mechanism is
by scattering with the conduction electrons via a single contact
interaction, such as in the alkali earth metals,  the Knight shift
and the \slrr\ are related \cite{korringa}: $T_1TK^2= \kappa $,
where the Korringa constant is given by: $\kappa= \mu_B^2/\pi k_B
\hbar \gamma^2$. For convenience, we define $S \equiv
T_1TK^2/\kappa$, so for the simple case, $S(T)$ is unity. Clearly
this quantity approaches a constant value at low temperatures, as
seen in the inset of Fig. (\ref{fig:T1Tdep}). This result suggests
the absence of magnetic correlations or magnetic order of the Pu
spins, which would likely contribute a strong temperature
dependence to $T_1T$.

The fact that $S(T\rightarrow0) \sim 0.05$ suggests that hyperfine
interaction in \dpu\ is more complex than in the alkali earth
metals. Electron-electron correlations can give rise to an
increase in $S(T)$, however $S>1$ is seen only in systems with a
single hyperfine coupling \cite{pines,CPSbook}. The fact that
$S(T)<1$ suggests that there is more than one hyperfine coupling
mechanism to the conduction electrons. In particular, if there
were two hyperfine couplings, $A_1$ and $A_2$, to two different
conduction bands, then the Knight shift would be given by the sum
$A_1+A_2$, whereas the \slrrtext\ would be given by the sum of the
squares $A_1^2+A_2^2$, so that naively one might expect $T_1TK^2
\sim \frac{A_1+A_2}{A_1^2+A_2^2}$.  If $A_1$ and $A_2$ have
different signs, then the measured ratio can be less than unity,
as we observe. In fact, the measured $S(T\rightarrow0)$ suggests
$A_2/A_1 = -0.9$. A second, negative hyperfine coupling via core
polarization is common in transition metals such as Platinum
\cite{clogston}. Given the complex electronic structure of \dpu,
and the result that the f electrons have itinerant behavior, such
a result is not surprising \cite{fradin,joyce}.

As seen in the inset, $S(T)$ has a quadratic temperature
dependence; we fit the data to $S(T) = S_0(1+(T/T_0)^2)$.  In
transition metals, $T_0$ is a measure of the energy scale on which
the density of states has structure away from the Fermi energy.
Specifically, Yafet and Jaccarino find $\pi k_BT_0=
\sqrt{\langle\rho/3d^2\rho/dE^2\rangle_{E_F}}$, where $\rho$ is
the density of states \cite{jaccarino}. Comparison with
photoemission data or band structure calculations is difficult,
however, since $k_BT_0 = 2.2$meV is less than their typical energy
resolution \cite{joyce,wills}.

\subsection{Inhomogeneous Broadening}
\begin{figure}
  \centering
  \includegraphics[width=\linewidth]{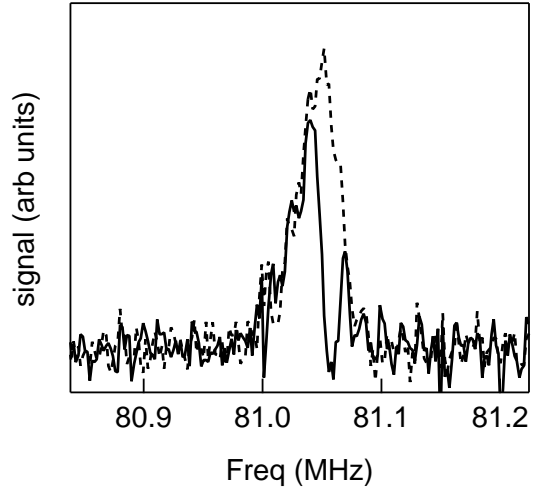}
  \caption{The hole-burning spectrum of $^{69}$Ga at 4K. The solid line is the spectrum after a narrow inversion pulse, and the
  dashed line is the spectrum with no preliminary pulse. }
  \label{fig:hole}
  \end{figure}

\begin{figure}
  \centering
  \includegraphics[width=\linewidth]{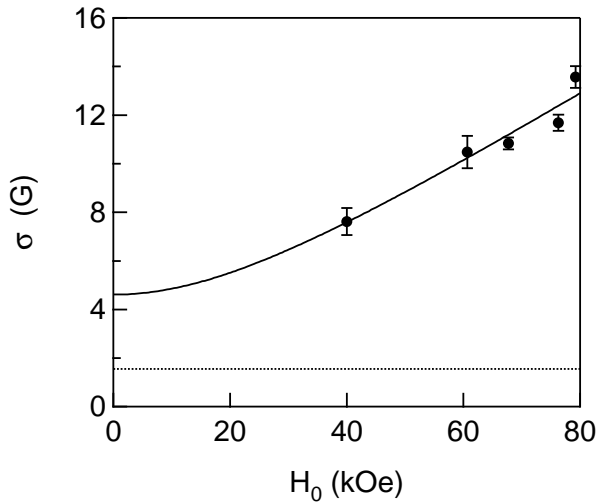}
  \caption{The rms second moment of the spectral line versus applied field. The solid line is a fit to Eq. (\ref{eqn:sigma}), and the
  dotted line is the estimated homogeneous linewidth. }
  \label{fig:FWHMvsFreq}
  \end{figure}

We now turn to the width of the NMR line.  In principle, the
resonance can be broadened by inhomogeneities in $H_0$, by
magnetic interactions with the other nuclei in the system, by
local magnetic moments on the Pu sites, and by quadrupolar
effects. The magnet has a homogeneity of 10ppm, so field
inhomogeneities can be neglected. Let us define $\sigma^2$ as the
second moment of the NMR line spectrum. For a nucleus with
$I>\frac{1}{2}$ we expect quite generally:
\begin{equation}
\sigma^2 = \sigma_{\rm mag}^2+\sigma_q^2 + (\chi_nH_0)^2 +
\sigma^2_{\eta\eta} + \sum_{\eta'\neq \eta} \sigma^2_{\eta\eta'},
\label{eqn:sigma}
\end{equation}
where $\sigma_{\rm mag}^2$ is the contribution from magnetic
moments in the system, $\sigma_q^2$ is the contribution from
non-zero EFG's at the nuclear site, $\chi_nH_0$ is the
contribution from the anisotropic susceptibility and
demagnetization fields in the powder, and $\sigma_{\eta\eta'}^2$
is the second moment of the nuclear dipole interaction. Note that
$\eta=\eta'$ corresponds to like spin coupling (for example
$^{69}$Ga-$^{69}$Ga coupling) and $\eta\neq\eta'$ corresponds to
unlike spin coupling ($^{69}$Ga-$^{71}$Ga and
$^{69}$Ga-$^{239}$Pu) \cite{CPSbook}.  These quantities can be
calculated for the fcc Pu lattice with randomly located Ga atoms,
and depend on the orientation of the field with respect to the
crystal parameters \cite{CPSbook}. For our polycrystalline sample,
the broadening is an average over the unit sphere; since
$\sigma_{\eta,\eta'}^{2}$ varies by less than 50\% over the unit
sphere we have chosen the (100) field direction for concreteness.
We find $\sigma_{69,69}^2=0.009 {\rm G}^2$,
$\sigma_{69,71}^2=0.004{\rm G}^2$, and $\sigma_{69,239}^2=2.5{\rm
G}^2$.  As seen in Fig. (\ref{fig:FWHM}), the spectral broadening
is much greater than the dipolar coupling.  In order to determine
the field dependent contribution, we measured the linewidth as a
function of applied field, as seen in Fig. (\ref{fig:FWHMvsFreq}).
Fitting the data to Eq. (\ref{eqn:sigma}) we find $\sigma(H_0=0) =
4.6$G, a value still greater than than dipolar second moment, so
the NMR spectrum must be inhomogeneously broadened by $\sim 4.3$G
$\sim 4.4$kHz.

To test for inhomogeneous broadening we performed a hole burning
test by saturating a narrow fraction of the line.  We applied a
low power pulse to the system prior to a broad-band echo sequence;
the spectrum is shown in Fig. (\ref{fig:hole}). Clearly, the full
spectrum consists of the superposition of several narrow intrinsic
lines.  Note that the rms second moment for the "hole" is 6.7G, a
value on the order of the field independent broadening, but still
greater than the calculated value.  However, the excitation pulse
in this case was still greater than the hole linewidth (but
narrower than tha full spectral linewidth) , so it is doubtful
that the measured hole linewidth reflects the intrinsic linewidth.

The inhomogeneous broadening arises either from magnetic moments
in the system, or from a distribution of EFGs.  In the latter
case, we utilize the measured hyperfine coupling  to estimate the
size of the putative magnetic moment as $5\times10^{-5}\mu_B$.
Such moments would not be detected in bulk susceptibility
measurements. However, neither $\sigma$ nor $K$ show any
significant temperature dependence, as one might expect if these
local moments were present in the system.

In fact, the most likely source of inhomogeneous broadening is
from a distribution of EFGs which contribute the the quadrupolar
linewidth.  One way to test for this is to compare the linewidth
of the $^{71}$Ga to that of the  $^{69}$Ga.  However, we find the
surprising result that the $^{71}$Ga width is 76\% larger than
that of the $^{69}$Ga, whereas the quadrupolar moment of the
$^{71}$Ga is 40\% \textit{smaller}!  (Note, however, that the
measured $\chi_n$'s (see Eq. \ref{eqn:sigma}) of the two isotopes
do follow the ratio of the gyromagnetic ratios, as expected.)  The
reason for this discrepancy may be related to the fact that the
sample had been held at temperatures $T<100$K for several days
prior to the measurements of the $^{71}$Ga. At these temperatures,
the damage to the lattice inflicted by the radioactive decay of
the Pu atoms is not annealed out, and is reflected in resistivity
measurements \cite{fluss}.  One expects that the EFG  will reflect
lattice damage, thus it is reasonable that the inhomogeneous
broadening is due to lattice distortions. In this case, the 4kHz
broadening would correspond to a distribution of $e^{69}QV_{cc}/h
\sim 27$kHz in the local EFG at the Ga site.

Any perturbation from cubic symmetry at the Ga site will give rise
to a finite EFG and consequently contribute to the line breadth.
XAFS studies of the Pu-Ga distances in \dpu\ indicate that the
lattice contracts slightly around the Ga impurities
\cite{conradson}, however it is not clear how the EFG will be
modified as a result. Further studies are needed to determine what
fraction of the quadrupolar line broadening arises from
radioactive damage and what arises from lattice contraction around
the Ga impurities.

\subsection{Decay of the Echo Envelope}

The decay of the echo envelope can provide information about the
like-spin coupling and the intrinsic linewidth.  The spin echo is
acquired by applying the pulse sequence $90^\circ - \tau -
180^\circ$, and occurs at a time $2\tau$ after the first pulse.
The integral of the spin echo is plotted versus $\tau^2$ in the
inset of Fig. (\ref{fig:FWHM}).  In the limit where
$\sigma^2_{\eta\eta}$ is highly anisotropic, so that the like-spin
nuclear coupling is much larger in a particular direction, the
form of the echo decay can be solved exactly in terms of
$\sigma^2_{\eta\eta}$ \cite{pennington,recchia,curroslichter}. In
this case, the echo decay is  Gaussian:
\begin{equation}
M(\tau)=M_0\exp\left({-\frac{(2\tau)^2}{2T_{2G}^2}}\right),
\label{eqn:edk}
\end{equation}
where $T_{2G}^{-2} = \gamma^2\sigma^2_{\eta\eta}$. However, for
the dipolar couplings discussed here, we are not in such a limit,
and one cannot write down an exact form for the echo decay
\cite{fine}. Nevertheless, we find that the echo decay is indeed
Gaussian with a time constant $T_{2G}\sim 480\mu$s. This value
corresponds to a second moment of 0.105G$^2$, a factor 10 times
larger than calculated. Even though we do not have an exact form
for the echo decay for dipolar couplings, one would expect a
priori that the measured echo decay constant would be within an a
factor of 2-3 of the like-spin second moment. Therefore, the fact
that we find such a disparity is surprising. If the like-spin
coupling is in fact larger than we estimated, then this result
suggests one of three causes: (i) there exist indirect couplings
between the Ga (unlikely due to their large spatial distances), or
(ii) the effective Ga-Ga distance is smaller, as might be expected
for Ga clustering, or (iii) the unlike Pu spins are fluctuating
quickly, so that there is a fluctuating field at the Ga site that
contributes to the dephasing of the Ga spins
\cite{currocouplingHighTc}. XAFS studies of the Ga distribution
suggest that the Ga is distributed uniformly, so the most likely
explanation for the enhanced echo decay rate is (iii)
\cite{conradson}.  In fact, we do expect that the Pu nuclear spins
have a very fast \slrrtext, since they must have a large hyperfine
coupling.

\subsection{Conclusions}

Ga NMR in \dpu\ provides information both about the local
structure and distribution of the Ga atoms, as well as the
electronic spin fluctuations.  We find that for 1.7 atomic percent
doping, \dpu\ shows little evidence for local magnetic moments at
the Pu sites, but that the hyperfine coupling between the Ga and
the conduction electrons probably contains a contact as well as a
core polarization term.  Furthermore, the NMR spectrum is
inhomogeneously broadened by a distribution of EFG's at the cubic
symmetric Ga site.  More detailed studies of the linewidth as a
function of doping should yield important information about the Ga
distribution.  It is worth noting that these experiments were
conducted at temperatures lower than the proposed Kondo
temperature of 200-300K \cite{fournier}.  Therefore, further
studies at higher temperatures may shed light on the presence of
spin fluctuations.

This work was performed at Los Alamos National Laboratory under
the auspices of the US Department of Energy. We thank J. Sarrao,
J. Thompson, J. Wills, M. Fluss, and C. P. Slichter for
enlightening discussions.


\begin{thebibliography}{99}

\bibitem{sarraoPuCoGa5nature} J. L. Sarrao, L. A. Morales, J. D. Thompson, B. L. Scott, G. R. Stewart,
 F. Wastin, J. Rebizant, P. Boulet, E. Collneau, and G. H. Lander,  Nature \textbf{420},
 297 (2002)


\bibitem{SmithBoringLANL}  A. M. Boring and J. L. Smith, Los
Alamos Science  \textbf{26} 90 (2000)


\bibitem{LashleySpecificHeat}  J.C. Lashley, J. Singleton, A. Migliori, J.B. Betts,
R. A. Fisher, J. L. Smith, and R.J. McQueeney,  Phys. Rev. Lett.
\textbf{91} 205901 (2003)

\bibitem{wills}  J. M. Wills \etal, cond-mat/0307767 and references
therein.

\bibitem{sarasov}  S. Y. Savrasov, G. Kotliar and E. Abrahams,
Nature \textbf{410} 793 (2001)

\bibitem{hecker}   S. S. Hecker, Los
Alamos Science  \textbf{26} 90 (2000)

\bibitem{magneticPutheory}  J. Bouchet, B. Siberchicot, F. Jollet and
A. Pasturel, J. Phys.: Condens. Matter \textbf{12} 1723 (2000)

\bibitem{conradson}  S. D. Conradson, Appl. Spec. \textbf{52} 252
1999)

\bibitem{Fradin1970}  F. Y. Fradin and M. B. Brodsky, Intern. J.
Magnetism \textbf{1} 89 (1970)

\bibitem{fournier}  S. Meot-Reymond, and J. M. Fournier,  Jour. Alloys and Compounds;
\textbf{232} 119 (1996)

\bibitem{korringa}  J. Korringa, Physica \textbf{16} 601 (1950)

\bibitem{pines}   D. Pines, Solid State Physics \textbf{1}, ed. by
F. Seitz and D. Turnbull (Academic, New York, 1955)

\bibitem{CPSbook}  C. P. Slichter, {\it Principles of Magnetic Resonance}, 3rd. Ed. (Springer
Verlag, New York, 1990)

\bibitem{clogston}  A. M. Clogston, V. Jaccarino, and Y. Yafet,
Phys. Rev. \textbf{134} 650 (1964)

\bibitem{joyce}  A. J. Arko, J. J. Joyce, L. Morales, J. Wills, J.
Lashley,F. Wastin and J. Rebizant, Phys. Rev. B. \textbf{62} 1773
(2000)

\bibitem{jaccarino}  Y. Yafet and V. Jaccarino, Phys. Rev.
\textbf{133} 1630 (1964)

\bibitem{fradin}  F. Y. Fradin in \textit{Plutonium and Other Actinides},
edited by H. Blank and R. Lindner, (North-Holland, Amsterdam,
1976)

\bibitem{fluss}  B. D. Wirth, A. J.  Schwartz, M. J. Fluss, M.J. Caturla, M. A. Wall, W. G. Wolfer,
MRS Bulletin \textbf{26} 679 (2001); M. J. Fluss, \etal, in press
J. Alloys Comp.


\bibitem{pennington}   C. H. Pennington and C. P. Slichter, Phys. Rev. Lett. {\bf 66}, 381 (1991)

\bibitem{recchia}  R. E. Walstedt and S- W. Cheong, Phys. Rev. B {\bf 51}, 3163 (1995);
C. H. Recchia, K. Gorny and C. H. Pennington, Phys. Rev. B {\bf 54}, 4207
(1996)

\bibitem{curroslichter}  N. J. Curro and C. P. Slichter,  J. Mag. Res. \textbf{ 130}, 186 (1998)

\bibitem{fine}  B. V. Fine, cond-mat/9707249

\bibitem{currocouplingHighTc}  N. J. Curro, J. Phys.  Chem.  Solids \textbf{ 63} 2181 (2002)

\end{thebibliography}
\end{document}